\documentclass[epj]{svjour}
\usepackage{latexsym}
\usepackage{graphicx}
\usepackage[numbers]{natbib} 
\usepackage{mhchem}
\usepackage{times,mathptmx}
\usepackage{balance} 
\usepackage{url}
\usepackage{lastpage}
\usepackage[format=plain,justification=raggedright,singlelinecheck=false,font=small,labelfont=bf,labelsep=space]{caption} 
\usepackage{fancyhdr}
\usepackage{color}  
\begin{document}
\title{Osmotic stress affects functional properties of human melanoma cell lines}
\author{Caterina A. M. La Porta\inst{1} \and Anna Ghilardi \and 
Maria Pasini\inst{1} \and Lasse Laurson\inst{2}\fnmsep \inst{3} \and Mikko J. Alava \inst{2} \fnmsep \inst{3} \and 
Stefano Zapperi\inst{3} \fnmsep \inst{4} \fnmsep \inst{5}\fnmsep \inst{6} \and Martine Ben Amar \inst{7}
\fnmsep \inst{8}
}                     
%
%
\institute{Department of Biosciences, University of Milano, via Celoria 26, 20133
Milano, Italy 
\and  COMP Centre of Excellence, 
Aalto University, P.O. Box 14100, FIN-00076 Aalto, Espoo, Finland
\and Department of Applied Physics, 
Aalto University, P.O. Box 14100, FIN-00076 Aalto, Espoo, Finland
\and Physics Department, University of Milano, via Celoria 16, 20133
Milano, Italy
\and CNR-IENI, Via R. Cozzi 53, 20125
Milano, Italy
\and ISI Foundation, Via Alassio 11C,
10126 Torino, Italy
\and Laboratoire de Physique Statistique, Ecole Normale Sup\'erieure, UPMC Univ Paris 06, Universit\'e Paris Diderot, CNRS, 24 rue Lhomond, 75005 Paris, France
\and Institut Universitaire de Canc\'erologie, Facult\'e de m\'edecine, Universit\'e Pierre et Marie Curie-Paris 6, 91 boulevard de l'h\^opital, 75013 Paris, France}
\date{Received: date / Revised version: date}

\abstract{Understanding the role of microenvironment in cancer growth and metastasis is a
key issue for cancer research. Here, we study the effect of osmotic pressure on the functional properties of primary and  metastatic melanoma cell lines. In particular, we experimentally quantify  individual cell motility and transmigration capability. We then perform a circular scratch assay to study how a cancer cell front  invades an empty space. Our results show that primary melanoma cells are sensitive to a low osmotic pressure, while metastatic cells are less.
To better understand the experimental results, we introduce and study a continuous model for
the dynamics of a cell layer and a stochastic discrete model for cell proliferation and diffusion.
The two models capture essential features of the experimental results and allow to make
predictions for a wide range of experimentally measurable parameters.
\PACS{
      {PACS-key}{discribing text of that key}   \and
      {PACS-key}{discribing text of that key}
     } 
} 
\maketitle

\section{Introduction}
The microenvironment is a key factor in tumour development and progression.  Its influence on
the behaviour of cancer cells is to a great extent mediated by the composition, structure, and dimensionality of the extracellular matrix, the polymeric scaffold that surrounds cells within tissues. Research shows that the mechanical properties of tumour microenvironment can facilitate or oppose tumour growth and dynamics, although this effect is poorly understood. Mechanical stresses such as those experienced by cancer cells during the expansion of the tumour against the stromal tissue have been shown to release and activate growth factors involved in the progression of cancer \cite{paszek2005}. Moreover, the stiffness of the matrix surrounding a tumour determines how cancer cells polarize, adhere, contract, and migrate, thus regulating their invasiveness \cite{paszek2005}. Another possibility is that mechanical stresses directly regulate the growth and death rates of cancer cells. Recently, Montel et al. have explored this possibility, investigating the effect of a constant stress applied on cellular spheroids over long time scales by inducing osmotic pressure by a solution of dextran, a biocompatible polymer which is neutral and can not be metabolised by mammalian cells \cite{montel2011,montel2012}. Using a similar method, some of us recently reported that a constant low osmotic pressure (1kPa) affects more strongly the proliferative capability of primary human melanoma cells (IgR39) in comparison to the corresponding metastatic ones (IgR37) \cite{Taloni2014}. Furthermore, a computer simulation analysis of the growth of melanocytic nevi inside epidermal or dermal tissues shows that the shape of the nevi is correlated with the elastic properties of the surrounding tissue\cite{Taloni2014}. 

Several studies in the literature reported important changes in cellular functioning due to osmotic pressure \cite{Racz2007,Nielsen2008}, but the stresses involved (in the MPa range) were orders of magnitude larger than those (in the kPa range) studied in Refs. \cite{montel2011,montel2012,Taloni2014}. It is interesting to notice that compressive stresses of slightly less than 1 kPa applied through a piston were recently found to induce a metastatic phenotype in cancer cells \cite{Tse2012}. In a recent paper, Simonsen et al. showed that interstitial fluid pressure (IFP) \cite{Fukumura2007} was associated with high geometric resistance to blood flow caused by tumour-line specific vascular abnormalities in xenografted tumours from two human melanoma lines with different angiogenic profiles \cite{simonsen2012}. In another recent paper, Wu et al. investigated how nonlinear interactions among the vascular and lymphatic networks and proliferating tumour cells might influence IFP transport of oxygen, and tumour progression \cite{wu2013}. They also investigated the possible consequences of tumour-associated pathologies such as elevated vascular hydraulic conductivities and decreased osmotic pressure differences. All these parameters might affect microenvironmental transport barriers, and the tumour invasive and metastatic potential, opening interesting new therapeutic perspectives.  The role of IFP for drug delivery was also analyzed computationally in Ref. \cite{Welter2013}. In general, understanding the influence of mechanical stress on cancer growth could shed new light on tumour development and progression.  Studies present in the  literature also use hypertonic conditions, but the effects are always observed at high osmotic pressure \cite{Bounedjah2012,Ignatova2006}. 

In this paper, we show that low osmotic pressure (1kPa) besides proliferation, which was the focus
of previous works \cite{montel2011,montel2012,Taloni2014}, also modulates key functional biological aspects of cancer cells such as their motility and transmigration, possibly contributing to the acquisition of a more aggressive phenotype. The choice of an osmotic pressure of 1kPa was suggested by previous experiments \cite{Taloni2014} showing that this is the smallest value that has an appreciable effect on the growth of melanoma cells. Melanoma is an aggressive, radio- and chemo-resistant tumour which becomes impossible to cure when metastasised. Therefore, a pressing clinical problem in the treatment of melanoma is to understand how to stop or prevent the capability of the tumour to give rise to metastasis. Using an interdisciplinary approach combining cellular biology and theoretical physics, we show that a low osmotic pressure (1kPa) leads to significant changes in cell F-actin organization compromising the capability of the cells to move and transmigrate. Interestingly, osmotic pressure is more effective on the primary human melanoma cell line with respect to the metastatic one belonging to the same patient.  We perform quantitative analysis on
the experimental data and extract parameters for cell motilities that we can then use in a continuum theory and a discrete model for cancer cell front dynamics. Overall, our results suggest that osmotic pressure could contribute to the selection of a subpopulation in the primary human melanoma cells contributing to the acquisition of a more aggressive phenotype. 

\section{Experimental results}

\subsection{Effect of osmotic pressure on F-actin organization}
\begin{figure}[ht]
\centering
\includegraphics[width=\columnwidth]{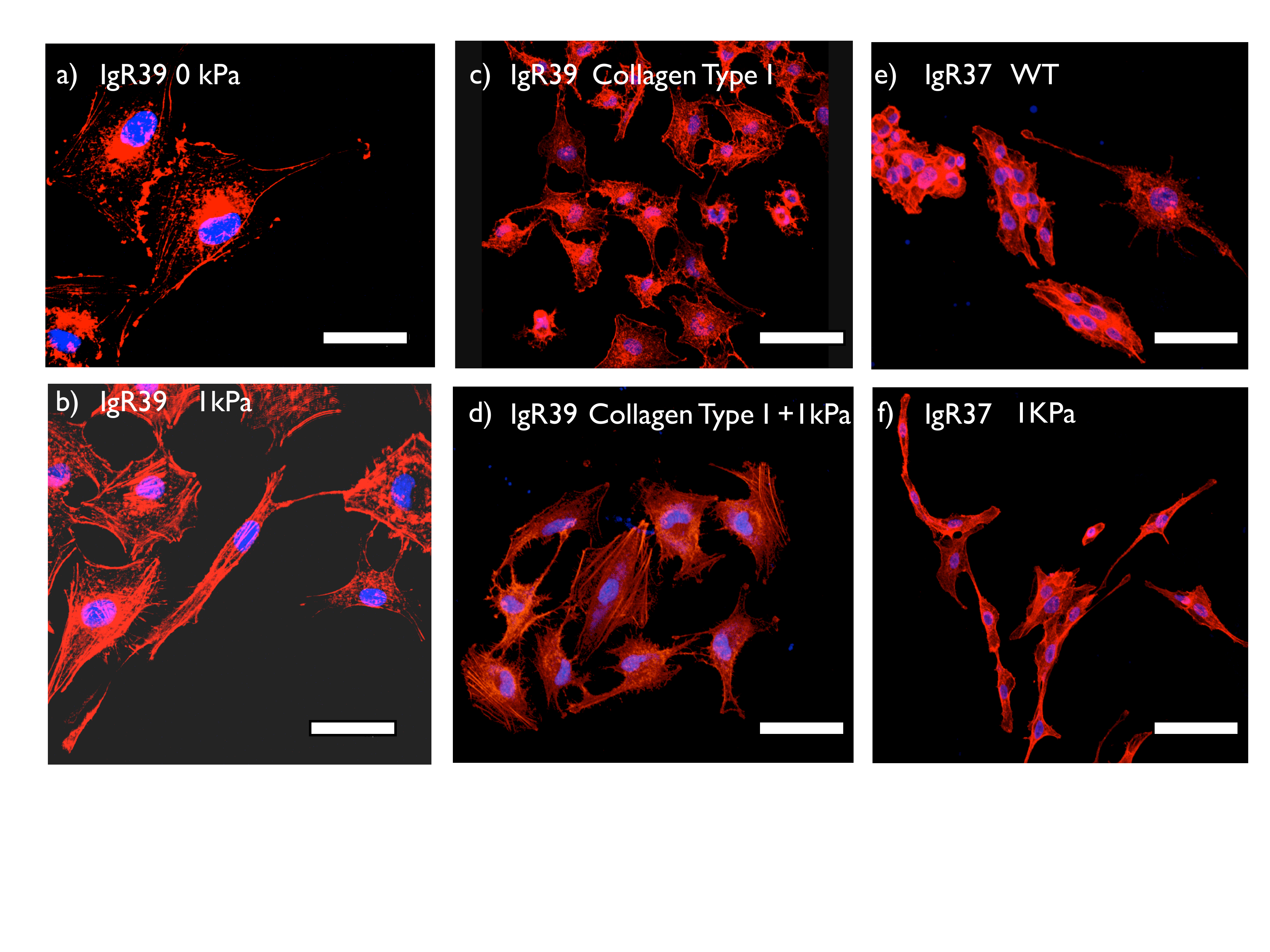}
\caption{{\bf Effect of osmotic pressure on F-actin.}
Subconfluent cells were plated on 33cm$^2$ Petri dish untreated or pre-coated with type I collagen. After 6 days the cells were fixed with 3.7\% paraformaldeide and incubated with Red-Falloidin for 30min at room temperature. The nuclei were counterstained with DAPI and the slides were mounted with Pro-long anti fade reagent (Invitrogen). The images were acquired using a Leica TCS NT confocal microscope. Panel a-d represent IgR39 cells and panel e-f IgR37 cells. Scale bar is 50$\mu$m }
\label{fig:1}
\end{figure}

The organization of F-actin in the two human melanoma cell lines is shown in Figure 1 under normal growth condition or 1kPa osmotic pressure without or with collagen as physiological substrate. In both cell lines, 1kPa induces a rearrangement in F-actin organization (Fig. \ref{fig:1}). In particular in IgR39, 1KPa osmotic pressure induces the appearance of filopodia and stress fibers (Fig. \ref{fig:1}a-b) Similar changes have been shown plating the cells on a collagen pre-coated Petri dish (Fig. \ref{fig:1}c-d). Indeed collagen should create a resistance against cell deformations that is similar to a confining pressure. The metastatic cell line IgR37 shows, however,  is only slightly affected by osmotic pressure displaying a more elongated shape (Fig.\ref{fig:1}e-f).

\begin{table*}
\begin{center}
\begin{tabular}{|c|c|c|c|c|}
\hline 
&	IgR39 0kPa &	IgR37 0kPa & IgR39 1kPa & IgR37 1kPa \\
\hline
without FBS	& $103\pm 14$	& $602\pm 74$	& $68\pm 11$	& $387 \pm 57$ \\
+FBS	 & $1704\pm 388$ &	$1925\pm 173$&	$269\pm 21$ &	$825 \pm 125$\\
\hline
\end{tabular}
\end{center}
\caption{\label{table:migration}  Result of the transwell assay for primary and metastatic human 
melanoma cells with and without osmotic pressure. The results are the mean number of cells crossing pores averaged  over three independent experiments. Error bars correspond to the standard error. }

\end{table*}

\begin{figure}[ht]
\centering
\includegraphics[width=\columnwidth]{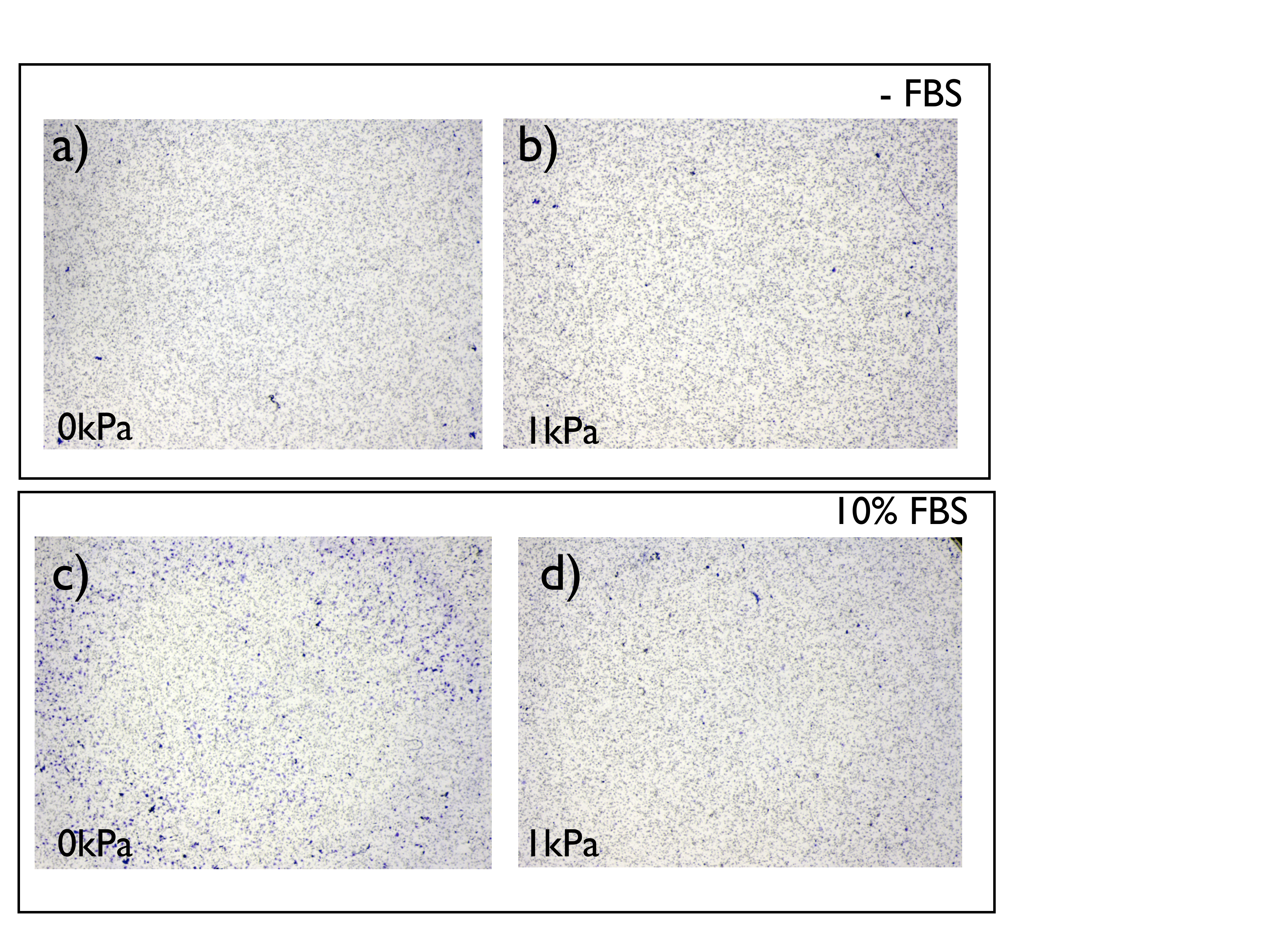}
\caption{{\bf Transwell Assay}. 70000 cells are plated to each transwell insert (8$\mu$m pores) into the upper chamber. In the lower chamber 600$\mu$l medium without FBS (-FBS) or when 10\% FBS is added. After 6 hours the cells are fixed and stained in a 20\% methanol/0.1\% crystal violet solution for three minutes at room temperature, followed by washing in deionized water to remove redundant staining. Non-migrated cells remaining at the upper side of the membranes, are carefully removed with cotton swabs and inserts are dried in darkness overnight. Pictures of the entire Petri dish are made with a Leica MZFL11 microscope mounted with a camera Leica DFC 32. The blue cells are counted using the magnified images (25X) and then calculated for the whole surface (32cm$^2$).}
\label{fig:6}
\end{figure}

\subsection{Effect of osmotic pressure on transmigration capability of IgR39 and IgR37 human melanoma cells}
An important function needed by  cancer cells in order to metastasise is the ability
to bypass an obstacle such as an endothelial wall. To simulate this process in vitro, we use a transwell assay counting the number of cells that are able to pass through 8$\mu$m wide pore after 6 hours in presence of a chemoattractant (10\%FBS). Without chemoattractant the cells barely overcome the obstacles, although metastatic cells are intrinsically more capable to do this  (Table \ref{table:migration}). This effect should, however, be considered just as background noise. Figure \ref{fig:6} shows a  significant decrease of transmigration capability for cells that are treated six days with a 1kPa osmotic pressure. We expose cells for some days to osmotic pressure in order to avoid 
measuring transient effects. The decrease is  by a factor 6 for IgR39 and by a factor 2  for IgR37 with respect to the untreated ones (Table \ref{table:migration}). 
These results show that osmotic pressure reduces the transmigration capability of melanoma cells but the effect is much stronger in primary IgR39 cells than in metastatic IgR37 cells.
\begin{figure}[htb]
\centering
\includegraphics[width=\columnwidth]{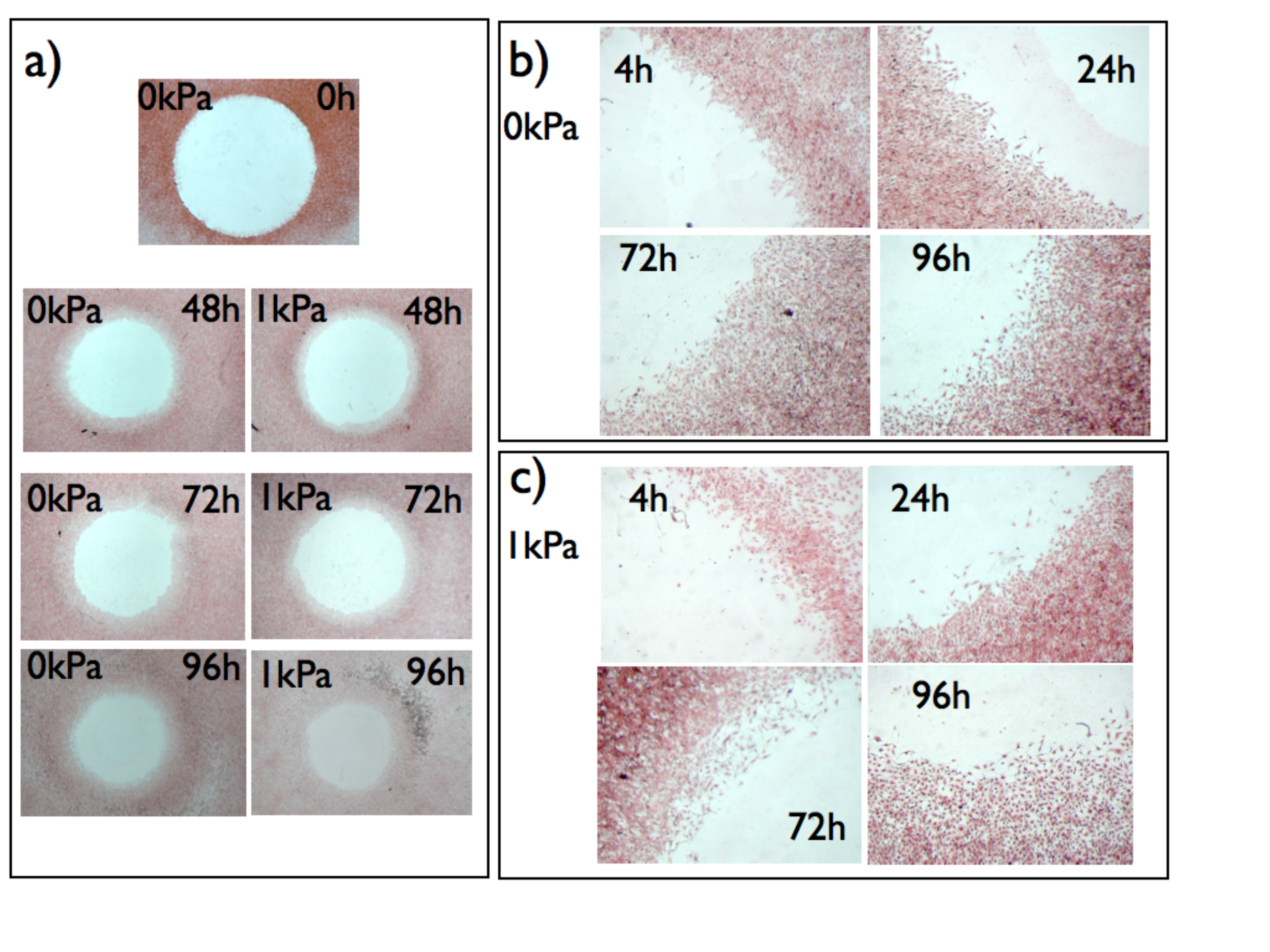}
\caption{{\bf Circular scratch assay}. Cells were plated on 33cm$^2$ Petri dish with at the center a disk (32mm$^2$) at confluence.The day after, the disk  was removed and cells were allowed to keep migrating and dividing for  different times, before they were fixed} with  3.7\% paraformaldeide and stained with hematoxillin/eosin solution at different times. The same experiment was carried out on cells treated with 1kPa osmotic pressure. Panel a shows a typical experiment at magnification 8X; Panel b-c show details from the same images at magnification 40X. The images were acquired with LEICA MZFLIII mounted with a camera LEICA DFC320. 
\label{fig:2}
\end{figure}

\subsection{Effect of osmotic pressure on circular scratch assay of IgR39 human melanoma cells.}

We study the biological effect of 1kPa osmotic pressure on the capability of the cells to 
cover a free space using a circular scratch assay.  As  in Fig. \ref{fig:2},  the circular shape of the disk is maintained during up to 96 hours, suggesting that the cells move isotropically into the free space. Moreover, in Table \ref{table:ring} is shown that osmotic pressure induces a little effect on the migration capability to cover the free area  (around 4\%). By observing the cells at higher magnification, we notice that the structure of the front is only slightly affected  by osmotic pressure\ref{fig:2}b-c.We also quantify the front roughness and find no significant difference for different osmotic pressures or times.
Colony formation assays reported in Ref. \cite{Taloni2014} for the same cells, show that the rate of division per day decreased from $k_{\mathrm{div}}=0.60$ to $k_{\mathrm{div}}=0.51$ under the effect of 1kPa of osmotic pressure. 
Here we investigate if the proliferative capacity of the cells are different at the border of the disk or at the edge of the Petri dish. As shown in Fig. \ref{fig:3}, no significant differences of proliferation have been shown using Ki67, a known proliferative marker, between the centre of the dish and the edge (Fig. \ref{fig:3} shows the expression at 72 hours, similar results was obtained at 24 and 48 hours). We did not carry out this kind of experiment in IgR37 since these cells do not grow as  a monolayer and it is impossible to get a homogeneous confluent layer. 
 \begin{table*}
\begin{tabular}{|c|c|c|c|}
\hline 
Time [hours] & Pressure [kPa] & Empty area [mm$^2$] & \% of area covered \\ 
\hline 
 0 & 0& $32.1 \pm 0.2$  &  $0 \pm 1$\\
 \hline 
24 & 0 & $21 \pm 2$  & $35 \pm 5$ \\ 
\hline 
24 & 1 & $23 \pm 5$ & $30 \pm 15$ \\ 
\hline 
48 & 0 & $16 \pm 1$  & $51 \pm 4$ \\ 
\hline 
48 & 1 & $21 \pm 4$ & $35 \pm 10$ \\ 
\hline 
72 & 0 & $17 \pm 1$ & $47 \pm 3$\\ 
\hline 
72 & 1 &  $14 \pm 5$ &  $57 \pm 16$\\ 
\hline 
96 & 0 &  $9 \pm 3$ & $70 \pm 9$\\ 
\hline 
96 & 1 &  $12 \pm 1$ &  $61 \pm 2$\\ 
\hline 
\end{tabular}
\caption{\label{table:ring} The table shows the free  area  
and the percentage of covered area obtained from Fig. \ref{fig:2}. Data are averaged
over two independent experiments and reported with the associated standard deviation.} 
\end{table*}
\begin{figure}[htb]
\centering
\includegraphics[width=\columnwidth]{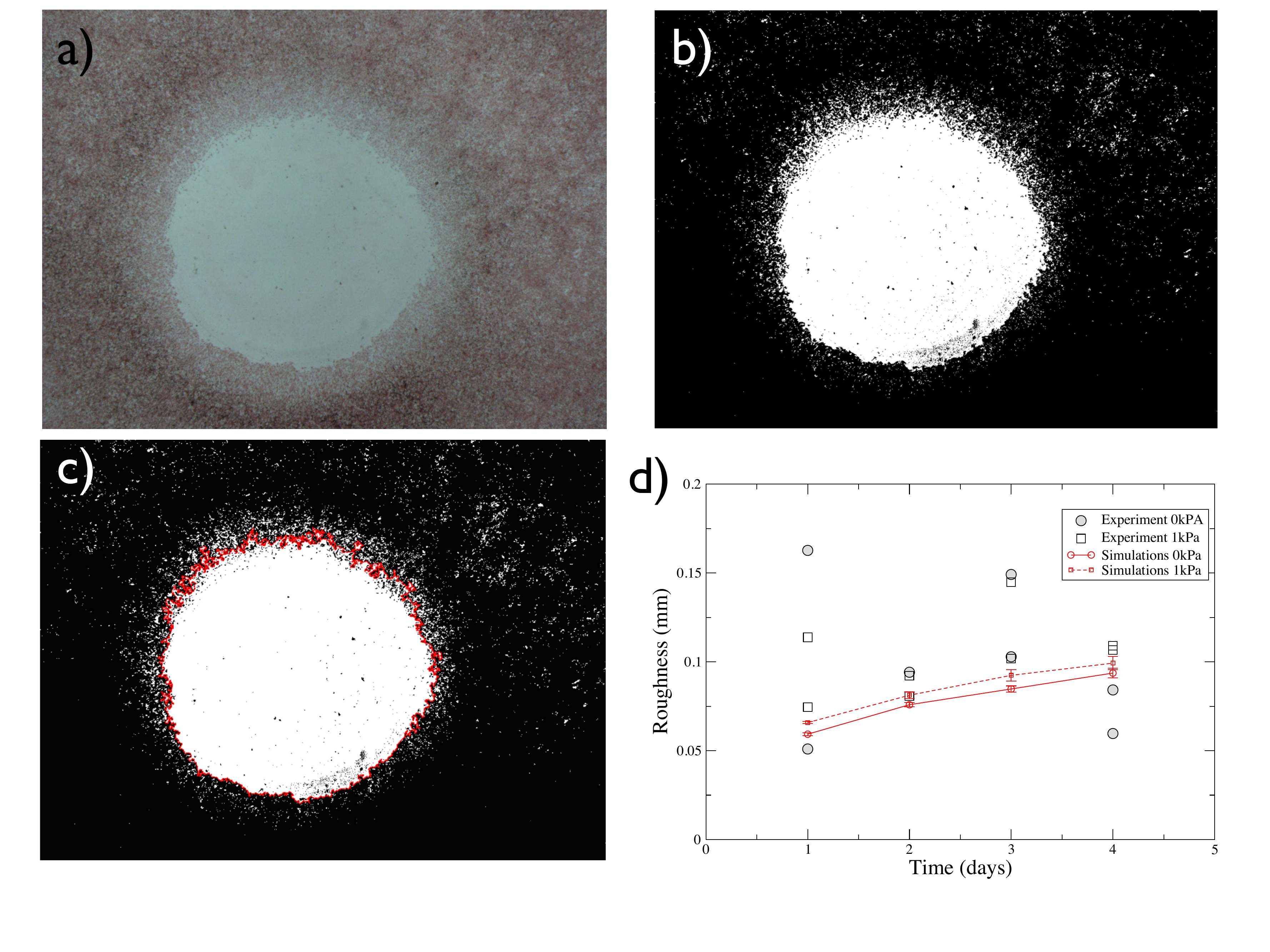}
\caption{
{\bf Estimate of the front roughness}. a) Plates resulting from the circular scratch assay are
digitized and b) thresholded using the same exposure and the same threshold for all the images. c) The cell 
front is obtained by a percolation cluster algorithm and d) the roughness of the front is computed for each
experimental condition (results of two independent experiments). 
For comparison, we also report the roughness obtained from the simulations of the computational model.}
\label{fig:roughness}
\end{figure}
\subsection{Osmotic pressure affects cell motility}
To investigate the effect of osmotic pressure on cell motility, we reconstruct cell trajectories
from confocal images (see \ref{fig:trajectories}a). We use IgR39 cells since these cells are
the ones that are most affected by osmotic pressure \cite{Taloni2014}. To quantify cell motility we compute the typical width of the trajectories $W= \langle (R(t) - \langle R \rangle)^2 \rangle$, where the average is taken over all time frames and 4 different cells for each experimental 
condition (see methods section). Fig. \ref{fig:trajectories}b shows that osmotic pressure reduces the average excursions of the cells but due to the large fluctuations the result is not statistically significant.To better differentiate between the two cases, we compute the time dependence of the mean square
displacement $ \langle (R(t+t') - R (t'))^2 \rangle$ which increases roughly linearly in
time (Fig. \ref{fig:trajectories}c) as would be expected for a Brownian motion. The slope
of the linear part of the curve can be used to estimate the diffusion coefficient resulting in
$D= 0.016 (\mu\mathrm{m})^2/\mathrm{s}$ at 0kPa and $D = 0.004 (\mu\mathrm{m})^2/\mathrm{s}$ at 1kPa. To confirm the Brownian nature of the process, we compute the power spectrum $S(\omega)$ of the trajectories, showing that it decays approximately as $1/\omega^2$ as expected for a random walk.
\begin{figure}[ht]
\centering
\includegraphics[width=\columnwidth]{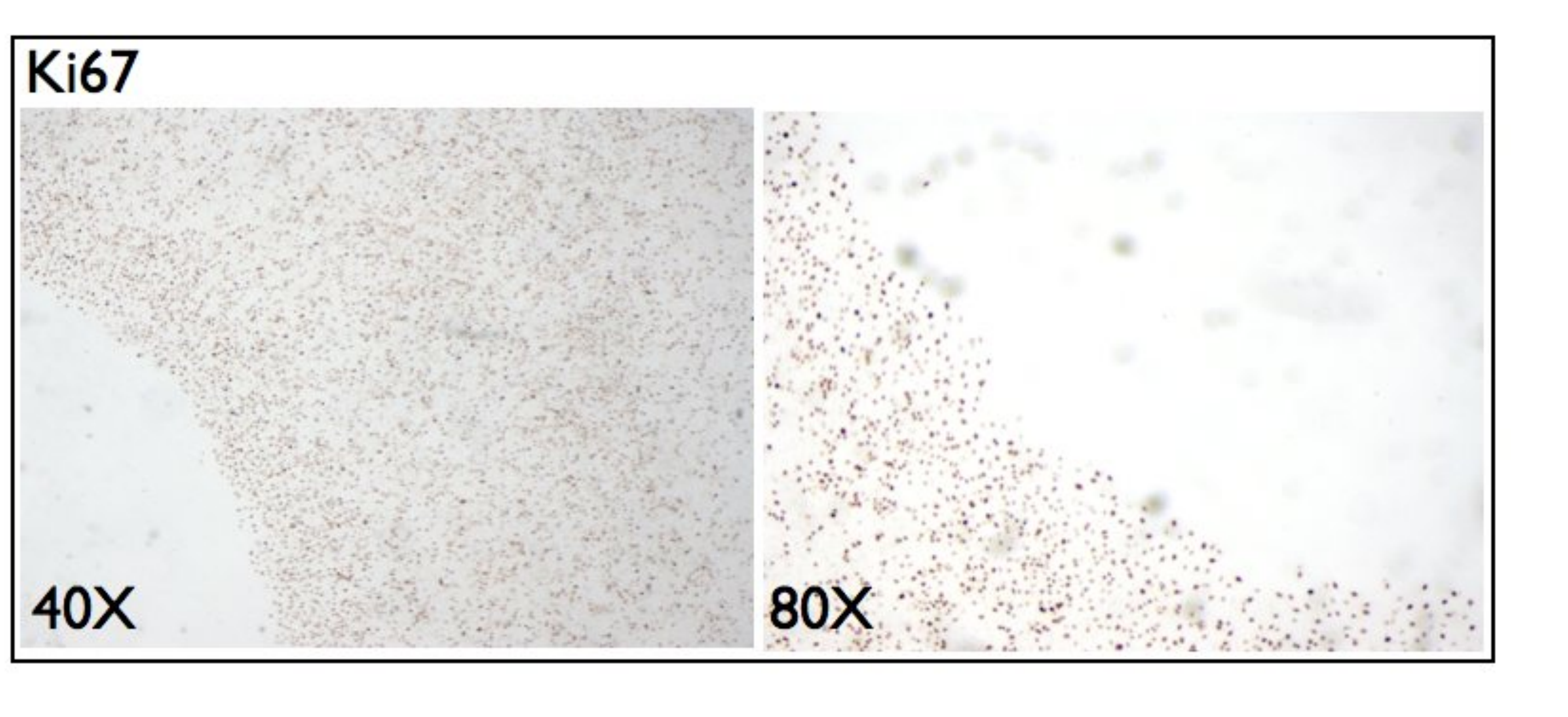}
\caption{{\bf Ki67 Immunohystochemistry}
IgR39 cells were plated on 33cm$^2$ Petri dish with at the center a disk (32mm$^2$) at confluence. 72 hours after the disk was removed, cells were fixed with 3.7\% paraformaldeide and stained overnight with anti-mouse Ki67 (1:100, DAKO). After a brief incubation for 30min with biotinylated anti-mouse secondary antibody(Dako, 1:100), the cells were incubated with streptavidin (Dako, 1:350) conjugated to peroxidase for 30min. Color was routinely developed using DAB peroxidase substrate kit (Vector) up to 10 min and coverslipped with a permanent mounting medium.  }
\label{fig:3}
\end{figure}
\begin{figure}[ht]
\centering
\includegraphics[width=\columnwidth]{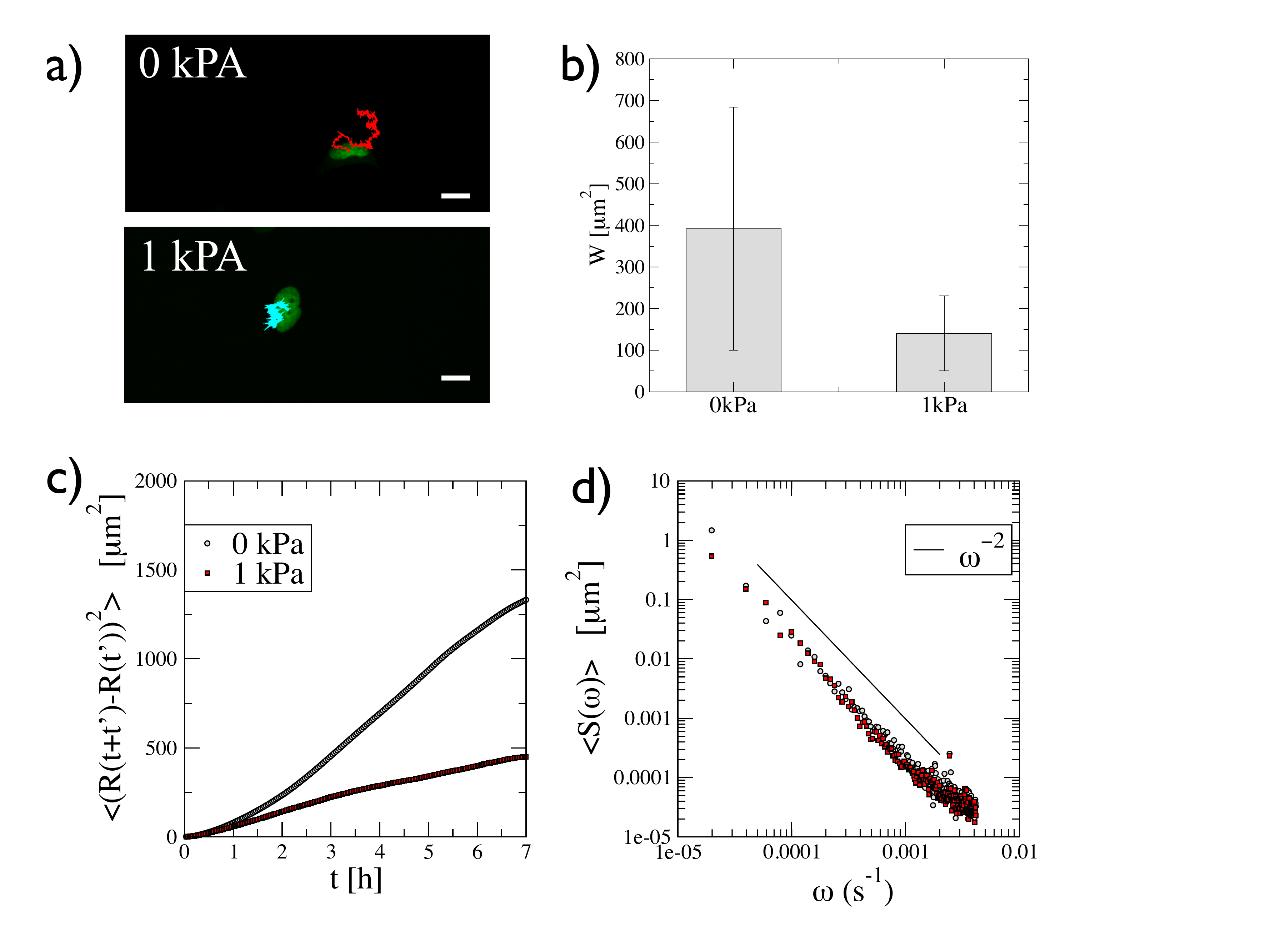}
\caption{{\bf Cell motility}. Subconfluent cells plated in a 35cm$^2$ Petri dish untreated or treated for 6 days with 1kPa of osmotic pressure are transfected with
H2B-GFP plasmide. Pictures are made on a confocal microscope (Leica TCS SP5 AOBS with resonant scanner, equipped with a 20X, 0.5 NA, dry objective) every 4 minutes for 14 hours.  a) Typical trajectories for IgR39 cells untreated and under osmotic pressure. b) The width of the trajectories 
is reduced under osmotic pressure but the result is not statistically significant due to the large error bars (n=4 for cells for each cases where considered). c) The average displacement grows roughly linearly with a typical slope that depends on the osmotic pressure. d) The Brownian nature of the motion is confirmed by the power spectrum of the trajectories that decays as $\omega^{-2}$. 
 }
\label{fig:trajectories}
\end{figure}
\section{A continuous model for advancing cell front}
Since the main objective of our work is to  study the effect of mechanical stress on an layer made of cancerous cells,  we propose a physical model to explain and quantify  the observations.  As shown in Table \ref{table:ring},  we get a strong decrease  of the area of the hole, about $50\%$, the first 2 days, then a slowest  decrease  as shown in Fig.(\ref{figosmo}). The effect is a little less acute if an over-pressure is added via addition of dextran. For an advancing layer under chemotaxis, a continuous model has already been proposed  \cite{Ben-Amar2013,Ben-Amar2014}, but these works are not concerned by a proliferative layer submitted eventually to an overpressure due to  an osmotic solution.  In addition, proliferation is  not often considered in the advance of a monolayer   but here our experiment clearly shows that it cannot be discarded.
Let us consider  a layer of constant thickness in a domain $\Omega$   with a border $\delta \Omega$. The cell displacement in $\Omega$ is mostly driven by the cell proliferation, but at the border the cells may be more  proliferative  eventually  and even  detach and move freely.  Due to proliferation, the mass conservation equation is then transformed  into:
\begin{equation}
\label{mass}
\frac{\partial \rho_0}{\partial t}+\vec \nabla  (\rho_0 \vec V)=-k_v (\tilde P-P_h),
\end{equation}
where $P_h$ is the homeostatic pressure for which cell proliferation just compensates  apoptosis, $\rho_0$ is the constant mass density,  $\vec V$ is the  velocity at an arbitrary point of $\Omega$.  This simple idea about  regulation  in living systems  is due to Claude Bernard, a french physiologist, in 1865 \cite{Bernard1865}. Mathematically, such a continuous model is valid at a scale larger than the cell size and the velocity is an averaged quantity on a sample of intermediate scale between the single cell and the full layer. Adding dextran to the solution increases  the pressure $\tilde P$,  slightly inhibiting the proliferation rate.  For a layer crawling on a substrate  in strong adhesion, the  second  Newton's law of dynamic simplifies into a Darcy law (\cite{Callan-Jones2008,Ben-Amar2013}) 
\begin{equation}
\label{Darcy}
\vec V=-\frac{1}{M} \nabla \tilde P,
\end{equation}
where the mobility coefficient $M$ is related to the  friction coefficient. Combining both equations gives:
\begin{equation}
\label{pressure}
\Delta P -\alpha^2 P=0\,\,\,{\mbox{with}}\,\,\, P=\tilde P-P_h
\end{equation}
and $\alpha^2=(k_v M {R_i^2})/\rho_0$, with $R_i$  the initial radius of  the hole of order the millimeter, chosen as our length unit in the following.
At the free interface, boundary conditions  include  the mechanical equilibrium and continuity of normal  velocities.
The pressure at the  layer  border $\tilde P_i$  is equal to the pressure of the solution (if  capillary effect is neglected). It is smaller at  the interface $\delta \Omega$  than in the layer bulk  $\Omega$ for a proliferative layer. The interface velocity $\vec{ \delta \Omega_t}$  is given by the velocity of the cells at the interface which may be different from the cells in the bulk $\Omega$. Indeed, these peripheral cells  have more room both for displacement and proliferation. Then the normal front velocity  $\vec{\delta \Omega_t} $  is given by
\begin{equation}
\label{velocity}
\vec N\cdot \vec{\delta \Omega_t}=-\frac{1}{M} \vec N\cdot  \vec \nabla  P + \vec N\cdot \vec v_s
\end{equation}
where $\vec v_s$ may represent a specific proliferation rate of the border cells.  
Other boundary conditions may exist due to the geometry of the experiment,  that is why we consider now $3$ different growth geometries. 
\subsection{Growing circular disc}
We imagine a disc of a cell monolayer and assume that the growth process preserves the circular geometry.  Eq.  (\ref{pressure}) for the pressure  $P=\tilde P-P_h$ is then transformed into an ordinary differential equation of second order:
\begin{equation}
\label{darcyrad}
\frac{1}{R}\frac{d}{dR} \big(R\frac{d P}{d R}\big)=\alpha^2 P
\end{equation}
The regular solution at the origin is the modified Bessel function of zero order $I_0$ so 
\begin{equation}
P(R)=P_i \frac{I_0 (\alpha R)}{I_0(\alpha R_i(t))}
\end{equation}
and   $P_i=\tilde P_i-P_h$ is negative, being fixed by the nutrient medium, whatever the  hole  radius $R_i(t)$. From Darcy's law (Eq. (\ref{Darcy})), we derive the 
growing velocity  at the interface:
\begin{equation}
\label{growthcirc}
\vec N\cdot \delta \Omega_t=\dot R_i(t)=-P_i \frac{\alpha}{M}  \frac{I_1 (\alpha R_i(t))}{I_0(\alpha R_i(t))}
\end{equation}
This equation means that  the interface velocity has the same velocity as the moving cells of the interface. 
As for the Saffman-Taylor viscous fingering in radial geometry \cite{Patterson}, our set of equations is  time independent but the time 
dependence reappears via the condition of continuity of velocities given by Eq.~(\ref{growthcirc}).  It is an implicit equation for the time variation of the radius which can be solved numerically. 
However one can make an asymptotic analysis of  Eq.(\ref{growthcirc}) and gets
 a constant growth rate at long times : 
 $$\dot R_i(t)\vert_{t \rightarrow \infty} \rightarrow   -P_i \frac{\alpha}{M} $$
  and a  radius growing linearly in time. One can notice  that a constant 
proliferation rate into Eq.(\ref{mass}) will lead to an exponential  time behavior for the radius so to follow the growth expansion of the layer at long times
may  be a good test for the validation of  the model.
\subsection{A hole in an infinite layer}
\label{referee}
In this case, we can accept the irregular Bessel function at the center $K_0$ 
 as the solution of Eq.(\ref{darcyrad}), but  not $I_0$, which grows exponentially. The
solution reads
\begin{equation}
\label{pression}
P(R)=P_i \frac{K_0(\alpha R)}{K_0(\alpha R_i(t))}.
\end{equation}
The pressure  $P$ vanishes at infinity where there is no cell proliferation.  The velocity given by Darcy's law is then:
 \begin{equation}
 \label{velib}
\dot R =\frac{\alpha}{M}P_i \frac{K_1(\alpha R)}{K_0(\alpha R_i(t))}
\end{equation}
being negative as $P_i$ and vanishing as $R \rightarrow \infty$  as expected.  Here again the radius of the hole is given numerically if we take Eq.(\ref{velocity}) on the interface.
In addition, the radius velocity diverges  when the total cover of the hole  is reached, and the continuous model stops to be valid for $R_i(t) \rightarrow 0$.  Indeed,
Eq.(\ref{velib}) is a nonlinear first order differential equation for the hole radius that we cannot integrate explicitly. However, for a large enough hole, assuming $\alpha$ of order $1$, the
 decrease  of the radius is linear  at initial time,  but the radius  will finally vanish like a square root, $(t-t_c)^{1/2}$, near the complete closure arising at time $t_c$.
Before presenting more elaborated model for our experiment, we would like to comment about the stability analysis of the  border and the assumption of circular symmetry for our set of equations.
 A careful observation of our  experiments presented in Fig. (\ref{fig:2}) shows that the inner border is somewhat noisy but does not present really instabilities similar to patterns observed in viscous flow \cite{Ben-Amar1993} or bacteria colonies \cite{Ben-Amar2013}.  It turns out that, even for a planar front, this model is stable against transverse perturbations, even in absence of capillary effect. Although astonishing, it is in agreement with observation. In order to convince the reader, we simply 
 present the   stability treatment  for this geometry in the Appendix section and analytically prove that our model predicts that the migration induced by growth  is stable in the circular geometry.  So the border irregularities observed in Fig. (\ref{fig:2})  are not fingering but can be interpreted as  the stochastic behavior of tumor cells.
 \begin{figure}[ht]
\centerline{\includegraphics[width=\columnwidth]{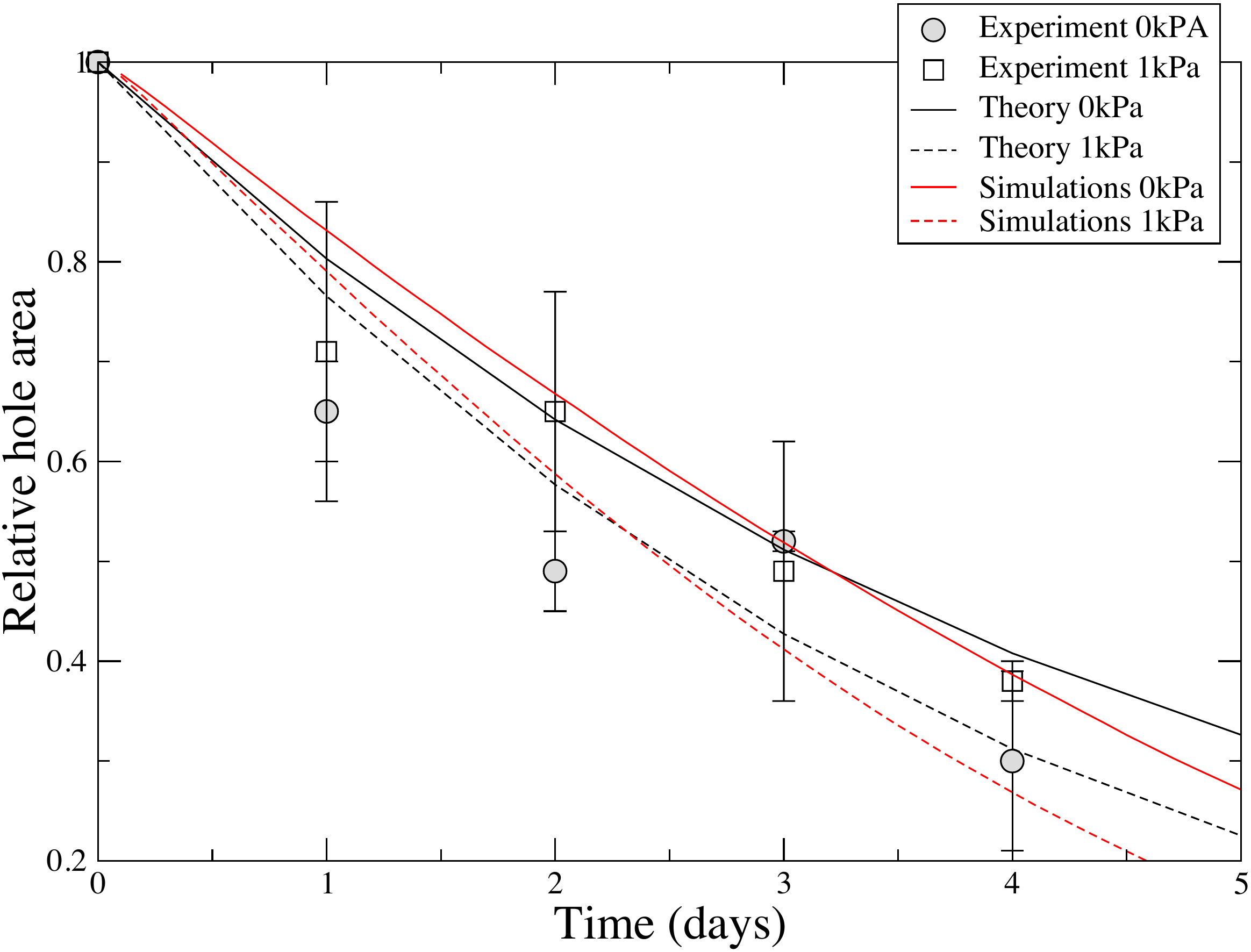}}
\caption{{\bf Front propagation.} Relative area of the hole as a function of time for experiments (see Table \ref{table:ring}), theory and simulations. Parameters for the theory are $P_i=-1$ for 0kPa and $P_i=-0.98$
for 1kPa, $\alpha_i=0.71,\alpha_e=0.53$, $\delta=0.3$ (with initial radius 1), $\tau=1.$ These numerical values correspond to the best estimated fitting  for our experimental data. Parameters for the
simulations are $k_{\mathrm{div}}=32$ division/day 
while keeping $k_{\mathrm{diff}}/k_{\mathrm{div}}=0.9$ or $1.7$, for 0kPa and 1kPa, respectively.}
 \label{figosmo}
\end{figure}
\subsection{The circular scratch assay}
Here we focus  on the geometry of the circular scratch assay: it is made of a planar annulus of cells limited by the rigid circular wall border of the Petri dish at  $R=R_e$ and , at $R=R_i$, is free to move into the empty space.
The interior interface at  $R=R_i$ is rather diffuse but an average circular border can be defined. In addition a thin circular ring (of thickness $\delta$)  with smaller cell density can be observed appearing less contrasted
than the inner core, located at  $R=R_i+\delta$. We will adapt the previous model to this inhomogeneous ring which has a density $\rho_i$ for  $R$ between $R_i(t)$ and $R_i(t)+\delta$ while the density in the bulk ($R_i(t)+\delta<R<R_e$)
 is called $\rho_e$ . It is important to remember that $\rho$ is indeed a density per unit surface, the volumetric density being the same for all cells. As a consequence, if we maintain the hypothesis of a monolayer, the surface density may be larger  at the inner border where the cells are more free to advance than in the annulus where they are more confined. It explains the optical difference in Fig. (\ref{fig:2}). We take constant  both $\rho$ values. Moreover, in vitro,  it has been shown  that in confluent layers, live cell extrusion is induced  to recover the homeostatic state  in case of  over-crowding and it is clear that the border allows 
 more easily such extrusion \cite{Eisenhoffer2012}. So we presume that $\rho_i<\rho_e$.
Because the layer geometry excludes $R=0$ and $R\rightarrow \infty$, possible solutions for Eq.(\ref{darcyrad}) are superpositions  of  $I_0$ and $K_0$ and we get 
\begin{equation}
\label{presi}
P^i(R)=\frac{P_i}{1+b} {\Big{\{}}\frac{I_0 (\alpha_i R)}{I_0 (\alpha_i R_i(t))}+b \frac{ K_0 (\alpha_i R)}{ K_0 (\alpha_i R_i(t))} {\Big{\}}}
 \end{equation}
with  $\alpha_i^2=k_v M/\rho_i$.
 The velocity of the cells at the outer border being :
\begin{equation}
\label{vitint}
\dot R^i=-\frac{\alpha_i}{M}\frac{ P_i }{1+b} {\Big{\{}}\frac{I_1 (\alpha_i R)}{I_0 (\alpha_i R_i(t))}-b \frac{ K_1 (\alpha_i R)}{ K_0 (\alpha_i R_i(t))} {\Big{\}}}
\end{equation}
$P_i$ keeps the same definition as before : the pressure in the solution minus the homeostatic pressure. For $R_i(t)+\delta<R<R_e$ we have:
\begin{equation}
\label{presb}
 P^e(R)=P_ e {\Big{\{}}\frac{ I_0 (\alpha_e R)}{I_1(\alpha_e R_e)}+ \frac{ K_0 (\alpha_e R)}{K_1(\alpha_e R_e)}{\Big{\}}}-\tau
   \end{equation}
$\tau$ being an evaporation rate of cells as mentioned by \cite{Eisenhoffer2012} and 
\begin{equation}
\label{vitb}
\dot R^e=-\frac{\alpha_e}{M} P_e  {\Big{\{}}\frac{ I_1 (\alpha_e R)}{I_1(\alpha_e R_e)}- \frac{K_1(\alpha_e R)}{K_1 (\alpha_eR_e)} {\Big{\}}}
\end{equation}
with $\alpha_e^2=(k_v M R_i^2)/\rho_e$. In Eqs.(\ref{presb},\ref{vitb}), we have applied the cancellation of cell velocity at the exterior border $R_e$.
We have $2$ unknowns: the coefficient $b$ and $P_e$ which are fixed  by the mechanical equilibrium $P^i(R_i(t)+\delta)= P^e(R_i(t)+\delta)$ and the continuity of the flux: $\rho_i d P^i/d R=\rho_e  d  P^e/d R$ for $R=R_i(t)+\delta$. The principle of such analysis is simple but requires also a numerical solution
that we present  in Fig.(\ref{figosmo}) in comparison with experimental data.  We have four unknown parameters $\alpha_i,\alpha_e$, $\tau$ and $P_i$. Varying parameters, the comparison with the experiment is mostly controlled by the quantity $\alpha P_i  /M = \sqrt{k_v/(\rho_i M)} R_i $ which is the inverse of a typical time. In the experiment the time scale is the day. 
Taking into account the experimental results shown in Fig.(\ref{figosmo}), the relative area variation of the hole after $2$ hours is $0.1$ for an osmotic pressure variation of  $1$ kPa.  It corresponds to a variation of the front velocity in international unit  of $7 10^{-9}$ m/s (evaluated from the experimental data of Fig.(\ref{figosmo})) which gives  the unknown value of $M$  of $10^{14}$ Pas/m${^2}$, using Eq(\ref{vitb}). Once we have this value we can derived easily the proliferation rate  $k_v/\rho_0$  of order $5 10^{-9}$ SI.  The friction coefficient $M$ for  an advancing epithelium  has been evaluated in \cite{Ben-Amar2013} for MDCK cells giving a value an order of magnitude larger $10^{15}$. Of course, the value of $M$ depends on the cells  involved in the experimental set-up, and are not  universal. However this estimation is consistent with usual order of magnitude  for biophysical values. 
 Taking again the order of magnitude of the velocity as $10^{-8}$m/s,  knowing that  forces one needs or can   apply to a cell are typically of order pico-Newton so $10^{-12}$ N  gives a friction coefficient of $ 10^{14}$ Ns/m$^4$ (being the ratio of  the force value by  cell volume and by  cell velocity). Finally sensitive variation by osmotic pressure being 0.02 in our model which corresponds to 1 kPa, gives a naive estimation of $50$ kPa for the departure from the homeostatic pressure. A better calculation can be done from the model using   Eq.(\ref{vitint}) and gives $25$ kPa.
\section{Computational model of cell duplication and movement}
\begin{figure}[ht]
\centering
\includegraphics[width=\columnwidth]{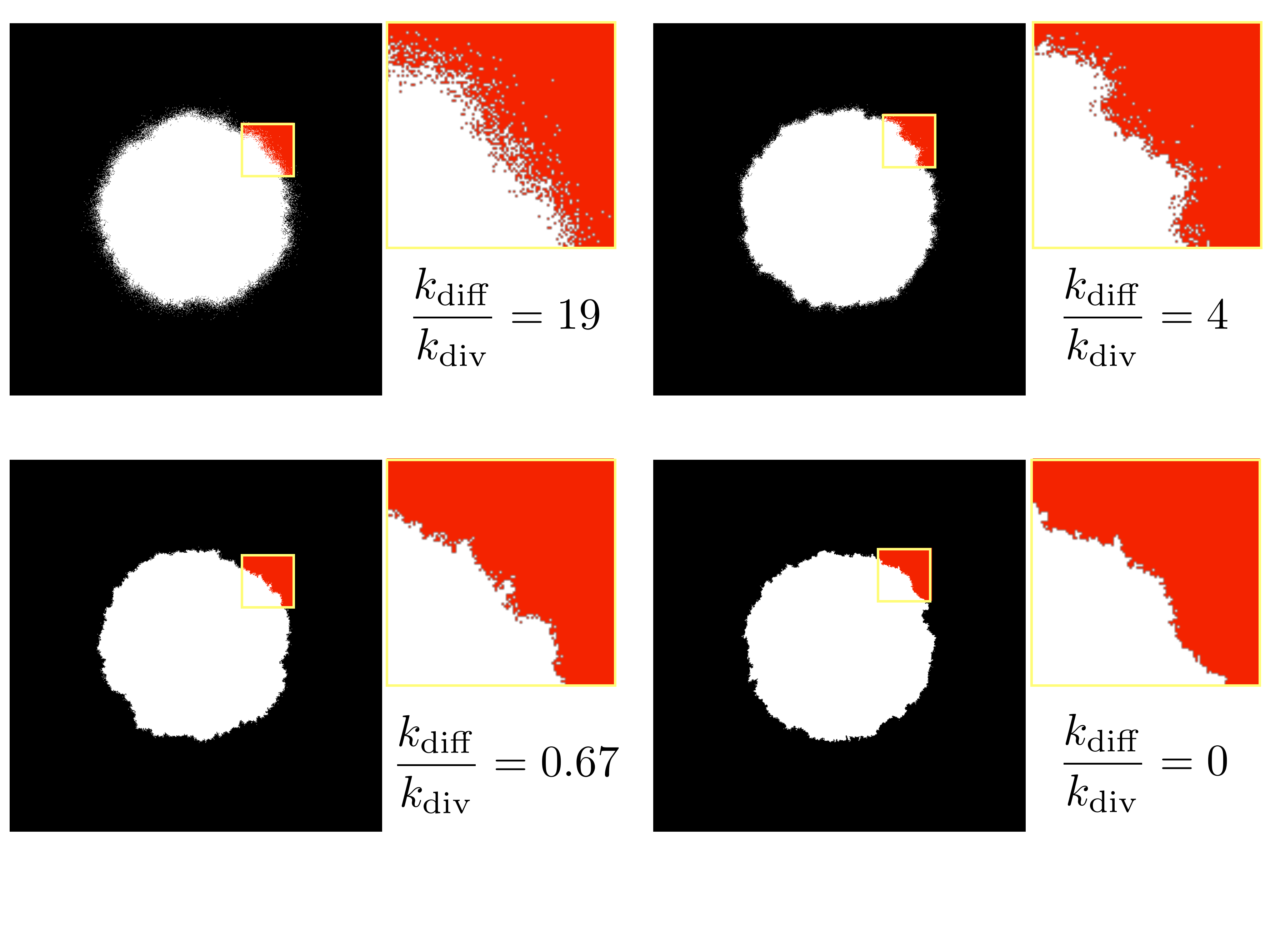}
\caption{{\bf Simulated ring assay}. Illustrative examples of the rings obtained for different ratios of the rates of diffusion and
division.}
\label{fig:4}
\end{figure}

\begin{figure}[ht]
\centering
\includegraphics[width=\columnwidth]{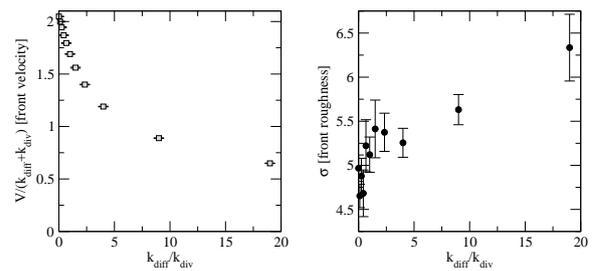}
\caption{{\bf Front velocity and roughness from numerical simulations.} a) The front velocity as a function of the ratio of the rates of diffusion and
division. b) The front roughness as a function of the ratio of the rates of diffusion and division.}
\label{fig:5}
\end{figure}

To understand how the morphology of the front in the circular scratch assay depends on the interplay between cell division and mobility, we simulate a simple lattice model including mechanisms for cell division and motility, with the relative rates of the two as the tunable parameter. The model is defined on a square lattice of linear size $L=700$, in which all the sites are occupied by cells except those inside a circle of radius $r=320$. Cells that have at least one unoccupied nearest neighbour site can either divide with rate $k_{\mathrm{div}}$ or diffuse with rate $k_{\mathrm{diff}}$.  The kinetics of the model follows the Gillespie algorithm \cite{gillespie1976}. At each time step we count how many cells have 1, 2, 3, or 4 neighbours (i.e. $n_1$, $n_2$, $n_3$ and $n_4$). Then the total rate is taken
to be $R_{\mathrm{tot}}=(n_4+0.75*n_3+0.5*n_2+0.25*n_1)*(k_{\mathrm{div}}+k_{\mathrm{diff}})$, where the reaction rate
is taken to be proportional to the number of neighbours a cell has, with the bare
rates given by those of a cell with all its four nearest neighbours vacant. 
At each time step, the reaction to take place is chosen randomly, with a probability
(rate of the reaction)/$R_{\mathrm{tot}}$. There are 8 possible reactions with different rates,
i.e. 4 possible division reactions and 4 possible diffusion reactions, each corresponding
to a different number of nearest neighbours a cell has (1,2,3 or 4). First, the
reaction to take place is chosen, and then a random cell with the number of nearest
neighbours corresponding to the selected reaction is found, and the reaction is
carried out. The  corresponding time step is generated from an exponential distribution with $R_{\mathrm{tot}}$ as the rate parameter. Then the simulation proceeds to the next time step, and the process continues until cells invade the center of the hole. Results are averaged over 10 realizations for each case.
In Fig. \ref{fig:4}, we report typical configurations of the circular scratch assay obtained by numerical simulations for different values of $k_{\mathrm{diff}}/k_{\mathrm{div}}$. When this ratio is large the interface is very diffuse, with several cells moving ahead of the front. As the ratio decreases, the front is more sharp, but still remains rough due to the randomness in the division process. In order to compare our simulations with experiments we first
estimate the ratio $k_{\mathrm{diff}}/k_{\mathrm{div}}$ from other experiments.
In particular, we can estimate $k_{\mathrm{diff}} \simeq D/a^2$, where $a$  is the
typical cell diameter. From Fig. \ref{fig:1}, we obtain that  $a\simeq 20\mu$m at 0kPa and 
$a\simeq 50\mu$m at 1kPa. Using our measurements for $D$ and $k_{\mathrm{div}}$ \cite{Taloni2014},  we obtain $k_{\mathrm{diff}}/k_{\mathrm{div}} \simeq 0.9$ at 0kPa and $k_{\mathrm{diff}}/k_{\mathrm{div}} \simeq 1.7$ at 1kPa.
If we use these value of $k_{\mathrm{div}}$ and $k_{\mathrm{diff}}$, the simulated front advances too slowly with respect to the experiments. This is expected since in the circular scratch assay cells are induced to rapidly
fill the hole. A good agreement between model and experiments is obtained setting $k_{\mathrm{div}}=32$ division/day 
while keeping $k_{\mathrm{diff}}/k_{\mathrm{div}}=0.9$ or $1.7$, for 0kPa and 1kPa, respectively.  The result is
reported in Fig. \ref{figosmo} for the front evolution and in Fig. \ref{fig:roughness} for the roughness.
The parameters employed in the model to obtain the best fit to the experiments look, however, unrealistic when compared
with estimates obtained from experiments on individual cells. One possibility could be the presence of {\it active} forces on
the cells that assists the closure of the hole. In the model, we have also quantified the front velocity and the roughness that are reported in Fig. \ref{fig:5}. The velocity is found to decrease when diffusion predominates over cell division, suggesting that the process of ring closure is dominated by cell division (Fig. \ref{fig:5}a).  Fig. \ref{fig:5}b shows instead that the front roughness increases when diffusion becomes more important. The effect of osmotic pressure is, however, very small and is hard to detect by  looking at the roughness, in agreement with experiments.
\section{Discussion}
Melanoma is one of the most aggressive tumours and practically impossible to cure once it becomes metastatic. Hence, one of the main goals of research is to understand the processes involved in metastasis. The classical approach is to consider the biological properties of the cell trying to identify biological markers. 
Here, we investigated this problem from a different point of view, focusing on the possible role of mechanical stress in the selection of a more aggressive cell subpopulation. Recent research has shown the involvement of mechanical stresses, such as the osmotic pressure, in the modulation of critical biological properties \cite{paszek2005,montel2011,montel2012,Taloni2014,Tse2012}. Therefore, we have investigated the effect of low osmotic pressure on two cell lines belonging to the same patient, IgR39 and IgR37 a primary and metastatic cell line, respectively. Our results show that a low osmotic pressure which gives little effect on cell proliferation on the primary human melanoma cell line \cite{Taloni2014}, induces significant changes on the F-actin organization,with the appearance of filopodia and stress fibres. These changes are more evident in primary human melanoma cell lines than in metastatic ones. Earlier studies on endothelial cells have shown that an osmotic pressure around 1.6 kPa induces an elongation of the cells due to a rearrangement of the cytoskeleton \cite{Acevedo1993,Salwen1998}, in agreement with our results.  In human breast cancer, mechanical compression has an effect  on the contractility of the cells suggesting the presence of a mechano-regulation which is able to integrate external physics stimulation into cytoskeletal changes and therefore cell phenotypes \cite{Tse2012}.
A critical question is whether F-actin rearrangement has a functional effect on biological properties of cancer cells such as their capability to move or to invade. Metastasis is indeed an integrated process composed by several steps. At least two key aspects play a critical role in this process: first the capability of the cell to move and then the capability of the cell to squeeze through the endothelium. To reproduce in vitro these two processes we have carried out the circular scratch assay which gives the possibility to study the movement of the cell into an open space and the transwell assay which investigates the capability of the cell to squeeze through a pore of a fixed size. Primary human melanoma cell are less able to move through fixed size pores using transwell assay in comparison to metastatic cells. Moreover, interestingly, low osmotic pressure leads to a reduced capability to transmigrate in both cell lines, however, the primary human melanoma cell line is more affected.  The circular scratch assay confirms an inhibitory effect of osmotic pressure on the capability to move of the cells. This is also confirmed by quantitative measurements of cell
motility from series of confocal images. 
In order to better understand the biophysical aspects of the circular scratch experiments, we have approached the problem theoretically and numerically. We have derived a continuum theory for the growth of a single layer of cells with different boundary conditions such those used in the circular scratch assay. This allows to obtain quantitative predictions for the dynamics of the cell front. In particular, we calculate the dependence of the front velocity on a few experimentally measurable parameters such as the cell division rate and motility.  The theoretical results are in good agreement with  the experiments. The main limitation of the theoretical model is that it does not include noise
and fluctuations which are obviously present in our experiments. To deal with this issue in a simple
way, we introduce a discrete lattice model of cell proliferation and diffusion which we simulate in
similar conditions. The discrete model allows to reproduce the experimentally observed roughening
of the cell front. This combination of theoretical and numerical models allows to understand better
the role of motility and cell division under osmotic pressure and provide a useful tool to explore 
the effect of experimentally relevant parameter. Both cell proliferation and motility rates are reduced by osmotic pressure, which also changes the shape of cells making them thinner and more elongated. The combination of this factor leads to an slight increase of  $k_{\mathrm{diff}}/k_{\mathrm{div}} $ under osmotic pressure.  
All together our results show how low osmotic pressure changes functional properties of  tumour cells but have less effect on more aggressive metastatic cell populations.  These results might be  in agreement with previous papers showing a role of osmotic pressure in the selection of more resistant subpopulation \cite{Boucher1991,paszek2005,Rofstad2002}.
\section{Material and Methods}
\subsection{Dextran solution}
A master solution of dextran at 10\% (w/v) was formed (Dextran from Leuconostoc spp, Fluka) and diluted to the desired concentration with complete medium. Transformation from dextran concentration to osmotic pressure was performed according to the calibration curve measured in Ref. \cite{bonnet-gonnet1994}.
\subsection{Cell lines}
Human IGR39 and IGR37 cells were obtained from Deutsche Sammlung von Mikroorganismen und Zellkulturen GmbH and cultured as previously described \cite{taghizadeh2010}. IGR39 was derived from a primary amelanotic cutaneous tumour and IGR37 from an inguinal lymph node metastasis in the same patient.
The cells were plated on pre-coated collagen type I plates according to the manufacturer's instructions for further experiments when specified in the results section.
\subsection{Cell transfection}
Cells were transfected with 1$\mu$g H2B-GFP plasmide (plasmide 116980, Addgene)  according to the manufactories instruction of SuperFect Transfection reagent by Qiagen. The cells were analyzed by time lapse using Leica TCS SP5 AOBS with resonant scanner, equipped with a 20X, 0.5 NA, dry objective (Leica Microsystems GmbH, Wetzlar Germany) with fluorescence under 5\% CO2 and 37C temperature.
\subsection{Immunofluorescence-Phalloidin}
Subconfluent cells grown on glass coverslip were fixed with 3.7\% parafolmaldeide in PBS for 10min, permealized with 0.5\% Triton X-100 in PBS for 5min at room temperature and. stained with 200$\mu$l of 100nM  Acti-stain\textsuperscript{\texttrademark} 555 Phalloidin (TebuBio, Cat. \# PHDH1) for 30min. DNA was counterstained for 30s with 200$\mu$l of 100nM DAPI and the slides were mounted with Pro-long anti fade reagent (Invitrogen).   The images were acquired using a Leica TCS NT confocal microscope.The same experiments were carried out using pre-treated collagen type I coverslips (Sigma). 
\subsection{Immunohystochemistry-Ki67}
Cells plated for the circular scratch assay  were fixed after 72 hours with 3.7\%paraformaldeide for 10min at room temperature, incubated for 1hr in 1\%BSA, 10NGS, 0.3M glycin, 0.1\%Tween in PBS and overnight with anti-Ki67 (clone MIB-1, Dako, 1:100 cod.) at 4C. All the sections were incubated for 30min with biotinylated anti-mouse secondary antibody in PBS (Dako, 1:100). This was followed by incubation with streptavidin (Dako, 1:350) coniugated to peroxidase in PBS for 30min, and by a brief rinse in PBS. Color was routinely developed using DAB peroxidase substrate kit (Vector)  up to 10 min, rinsed in distilled water, and coverslipped with a permanent mounting medium.  
The images were acquired using a microscope Leica MZ FLIII mounted with a camera Leica DFC320.
\subsection{Circular scratch assay}
Cells were plated on 33cm$^2$ Petri dish with a disk at the center  (Sigma cod. z370789, diameter: 6.4mm, area 32mm$^2$) to reach ~70-80\% confluence as a monolayer. When they were at confluence, the disk was removed. At different times  after removal (from 4 to 96 hours), the cells were fixed  with 3.7\% paraformaldeide a different times (from 4 to 96hrs) and stained with hematoxillin-eosin. The images were acquired using a Leica MZFL11 microscope mounted with a Camera Leica DFC 32 at different magnification (8, 16 and 25X). To quantify the free area and the correspondent percentage of covered area,  we used Gimp (GNU Image Manipulation Program) to threshold the images. The front profile is obtained by using a simple cluster algorithm, as illustrated in Fig. \ref{fig:roughness}. We then compute the 
area enclosed by the front and average the result over two independent realizations of the experiment under the same experimental conditions. We also use the same algorithm to obtain the front roughness $\sigma$, defined as
the standard deviation of the front radii.
\subsection{Transwell migration assay}
Migration experiments were conducted using a conventional 24-well Transwell system (6.5 mm TranswellH (\# 3422), CorningH, NY, USA) with each well separated by a microporous polycarbonate membrane (10 $\mu$m thickness, 8 $\mu$m pores) into an upper (insert) and a lower chamber (well) according to [16]. After 24 hours of serum deprivation, cells were detached, counted and resuspended in growth media without FBS to obtain equal cell densities (296000 cells/cm$^2$). A volume of 250 $\mu$l containing 70000 cells was plated to each insert and 600 $\mu$l medium was added to the wells. As chemotactic factors, a medium containing 10\% FBS was used. After 6 hours the cells were fixed and stained in a 20\% methanol/0.1\% crystal violet solution for three minutes at room temperature, followed by washing in deionized water to remove redundant staining [16]. Non-migrated cells remaining at the upper side of the membranes were carefully removed with cotton swabs and inserts were dried in darkness overnight.  The following day at stained membranes were made pictures using a transmitted-light microscope (Leica MZFL11 microscope mounted with a camera Leica DFC 32) at different magnification (8, 16 and 25X). The blue cells were counted using the magnified images (25X) and then calculated for the whole surface (32cm$^2$).

\subsection{Cell tracking}
Cell tracking has been done in ImageJ by first projecting the confocal stacks (maximum intensity) and then finding the maxima for different frames. {In each frame $i=1,...,N$, we identify the cell center of mass $(X_i,Y_i)$ and then compute the width of the trajectory as $W=\sum_{i} ((X_i-\bar{X})^2+ (Y_i-\bar{Y})^2)/N$,
where $\bar{X}=\sum_{i} X_i/N$ and $\bar{Y}=\sum_{i} Y_i/N$. The mean-square displacement is calculated
as
\begin{equation}
C(i) = \sum_{j=1}^{N/2} 2((X_{i+j}-X_{j})^2+(Y_{i+j}-Y_{j})^2)/N,
\end{equation}
and the power spectrum by taking the discrete Fourier transform of $(X_i,Y_i)$ and then squaring the amplitudes.

\subsection{Statistical analysis}
Statistical significance was evaluated using the Kolomogorov-Smirnov non parametric test (see http://www.physics.csbsju.edu/stats/KS-test.html).

\subsection{Appendix}
Let us consider  the stability analysis of the model of section \ref{referee}. For that we expand the pressure field and the inner border as
\begin{equation}
\tilde P(R,\theta,t)= P(R) +\epsilon p(R) \cos(m \theta) e^{\Omega_m t}
\end{equation}
\begin{equation}
\tilde R_i(\theta,t)=R_i(t)  +\epsilon \cos(m \theta )e^{\Omega_m t}
\end{equation}
where $P(R)$, the pressure calculated in the circular geometry and the inner hole radius  $R_i(t)$ are given respectively by  Eq.(\ref{pression},\ref{velib}). Hereafter we simplify the notations into $R_i$. 
Replacing $P$ by $\tilde P$ into Eq.(\ref{pressure}) gives for $p(R)$ the following equation:
\begin{equation}
R^2 p" + Rp' -(\alpha^2 R^2+m^2) p = 0
\end{equation}
whose regular solution at infinity is  the modified Bessel function of $m$ order  $K_m(\alpha R)$:
\begin{equation}
p(R)=p_0 \frac{K_m(\alpha R)}{K_m(\alpha R_i)}
\end{equation}
Because $\tilde P=P_i$ at $R_i$, the linear variation of the pressure vanishes at $R=R_i$ giving $p_0+P'(R_i)=0$. 
So  we deduce  the value of  $p_0=\alpha P_i K_1(\alpha R_i)/K_0(\alpha  R_i)$.
The continuity equation modified by the  border distortion is then
\begin{equation}
\Omega_m=-\frac{1}{M}( P"+p')
\end{equation}
After some elementary calculation involving the modified Bessel functions we get
\begin{equation}
\Omega_m=-\frac{\alpha^2P_i}{M} \Big{\{{}}1+\frac{K_1(\alpha R_i)}{K_0(\alpha R_i)}\Big(\frac{1-m}{\alpha R_i} -\frac{K_{m-1}(\alpha R_i)}{K_{m}(\alpha R_i)}\Big)\Big{\}}
\end{equation}
Taking the 2 extreme limiting values $R_i\rightarrow0$ or $R_i\rightarrowÊ\infty$, we find $\Omega_m$ always negative. Indeed for $R_i$ going to zero, the second term in the bracket is dominant and negative $(P_i<0)$
 \begin{equation}
 \Omega_m\sim -\frac{P_i(1-m^2)}{2M}
 \end{equation}
 which shows that $\Omega_m$ is always negative when the hole radius decreases. This stability linear analysis proves that,  for
  the circular geometry of the hole closing,  induced by cell  proliferation, our modeling  predicts a stable process. Although the equations have a strong similarities with Laplacian growth \cite{Ben-Amar1993}
  this model  predicts that there is no  fingering as time goes on.
  
  \section{Acknowledgments}
This work is supported in part by AAP Physique and Cancer 2012, project DERMA (MBA). SZ acknowledges the visiting professor programm of UPMC, the European Research Council Advanced grant SIZEFFECTS and the Academy of Finland FiDiPro progam, project 13282993.   CAMLP acknowledges the visiting professor programme of Aalto University and support from PRIN 2010. LL and MJA are supported by the Academy of Finland through an Academy Research Fellowship (LL., Project No. 268302) and through the Centres of Excellence Program (MJA and L. L., Project No. 251748).


\begin{thebibliography}{25}

\bibitem{paszek2005}
M.J. Paszek, N.~Zahir, K.R. Johnson, J.N. Lakins, G.I. Rozenberg, A.~Gefen,
  C.A. Reinhart-King, S.S. Margulies, M.~Dembo, D.~Boettiger et~al., Cancer
  Cell \textbf{8}(3), 241 (2005)

\bibitem{montel2011}
F.~Montel, M.~Delarue, J.~Elgeti, L.~Malaquin, M.~Basan, T.~Risler, B.~Cabane,
  D.~Vignjevic, J.~Prost, G.~Cappello et~al., Phys. Rev. Lett. \textbf{107},
  188102 (2011)
\bibitem{montel2012}
F.~Montel, M.~Delarue, J.~Elgeti, D.~Vignjevic, G.~Cappello, J.~Prost, New
  Journal of Physics \textbf{14}(5), 055008 (2012)

\bibitem{Taloni2014}
A.~Taloni, A.A. Alemi, E.~Ciusani, J.P. Sethna, S.~Zapperi, C.A.M. La~Porta,
  PLoS One \textbf{9}(4), e94229 (2014)

\bibitem{Racz2007}
B.~Racz, D.~Reglodi, B.~Fodor, B.~Gasz, A.~Lubics, F.~Gallyas, Jr, E.~Roth,
  B.~Borsiczky, Bone \textbf{40}(6), 1536 (2007)

\bibitem{Nielsen2008}
M.B. Nielsen, S.T. Christensen, E.K. Hoffmann, Am J Physiol Cell Physiol
  \textbf{294}(4), C1046 (2008)

\bibitem{Tse2012}
J.M. Tse, G.~Cheng, J.A. Tyrrell, S.A. Wilcox-Adelman, Y.~Boucher, R.K. Jain,
  L.L. Munn, Proc Natl Acad Sci U S A \textbf{109}(3), 911 (2012)

\bibitem{Fukumura2007} D. Fukumura,  R.K. Jain, J. Cell. Biochem. \textbf{101}, 937 (2007)

\bibitem{simonsen2012}
T.G. Simonsen, J.V. Gaustad, M.N. Leinaas, E.K. Rofstad, PLoS One
  \textbf{7}(6), e40006 (2012)

\bibitem{wu2013}
M.~Wu, H.B. Frieboes, S.R. McDougall, M.A.J. Chaplain, V.~Cristini,
  J.~Lowengrub, J Theor Biol \textbf{320}, 131 (2013)

\bibitem{Welter2013}
M. Welter and H. Rieger PLoS ONE \textbf{8}, e70395 (2013) 

\bibitem{Bounedjah2012}
O.~Bounedjah, L.~Hamon, P.~Savarin, B.~Desforges, P.A. Curmi, D.~Pastr{\'e}, J
  Biol Chem \textbf{287}(4), 2446 (2012)

\bibitem{Ignatova2006}
Z.~Ignatova, L.M. Gierasch, Proc Natl Acad Sci U S A \textbf{103}(36), 13357
  (2006)

\bibitem{Ben-Amar2013}
M.~Ben~Amar, Eur Phys J E Soft Matter \textbf{36}(6), 64 (2013)

\bibitem{Ben-Amar2014}
M.~Ben~Amar, M.~Wu, J R Soc Interface \textbf{11}(93), 20131038 (2014)

\bibitem{Bernard1865}
C.~Bernard, \emph{Introduction \`{a} l' \'{e}tude de la m\'{e}decine
  exp\'{e}rimentale} (Paris, 1865)

\bibitem{Callan-Jones2008}
A.C. Callan-Jones, J.F. Joanny, J.~Prost, Phys Rev Lett \textbf{100}(25),
  258106 (2008)

\bibitem{Patterson}
L.~Patterson, Journ. of Fluid Mech. \textbf{113}, 513 (1981)

\bibitem{Ben-Amar1993}
M.B. Amar, J. Phys.I France \textbf{2}, 353 (1993)

\bibitem{Eisenhoffer2012}
G.T. Eisenhoffer, P.D. Loftus, M.~Yoshigi, H.~Otsuna, C.B. Chien, P.A. Morcos,
  J.~Rosenblatt, Nature \textbf{484}(7395), 546 (2012)

\bibitem{gillespie1976}
D.T. Gillespie, Journal of Computational Physics \textbf{22}(4), 403  (1976),
  I

\bibitem{Acevedo1993}
A.D. Acevedo, S.S. Bowser, M.E. Gerritsen, R.~Bizios, J Cell Physiol
  \textbf{157}(3), 603 (1993)

\bibitem{Salwen1998}
S.A. Salwen, D.H. Szarowski, J.N. Turner, R.~Bizios, Med Biol Eng Comput
  \textbf{36}(4), 520 (1998)

\bibitem{Boucher1991}
Y.~Boucher, J.M. Kirkwood, D.~Opacic, M.~Desantis, R.K. Jain, Cancer Res
  \textbf{51}(24), 6691 (1991)

\bibitem{Rofstad2002}
E.K. Rofstad, S.H. Tunheim, B.~Mathiesen, B.A. Graff, E.F. Hals{\o}r,
  K.~Nilsen, K.~Galappathi, Cancer Res \textbf{62}(3), 661 (2002)

\bibitem{bonnet-gonnet1994}
C.~Bonnet-Gonnet, L.~Belloni, B.~Cabane, Langmuir \textbf{10}(11), 4012 (1994),
  
\bibitem{taghizadeh2010}
R.~Taghizadeh, M.~Noh, Y.H. Huh, E.~Ciusani, L.~Sigalotti, M.~Maio, B.~Arosio,
  M.R. Nicotra, P.~Natali, J.L. Sherley et~al., PLoS One \textbf{5}(12), e15183
  (2010)

\end{thebibliography}

\end{document}